\documentclass[prc,twocolumn,showpacs,floatfix,nofootinbib,preprintnumbers,superscriptaddress,amsmath,amssymb]{revtex4-1}

\usepackage{graphicx}
\usepackage{epsfig}
\usepackage{amsmath,amssymb}
\usepackage{bm}
\usepackage{color}
\usepackage{float}
\usepackage{dcolumn} 
\usepackage{enumerate}
\usepackage{multirow}
\usepackage{threeparttable}

\begin{document}

\newcommand{\pr}[1]{{\sc{\lowercase{#1}}}}
\newcommand{\bra}[1]{\langle #1 \vert}
\newcommand{\brared}[1]{\langle #1 ||}
\newcommand{\ee}{\eta^{\ast},\eta}
\newcommand{\product}[2]{\left\langle #1 | #2 \right\rangle}
\newcommand{\kbar}{\bar{k}}
\newcommand{\ket}[1]{\vert #1 \rangle}
\newcommand{\ketred}[1]{|| #1 \rangle}
\newcommand{\ahat}{\hat{a}}
\newcommand{\adag}{a^{\dagger}}
\newcommand{\ahatdag}{\hat{a}^{\dagger}}
\newcommand{\Ahat}{\hat{A}}
\newcommand{\Adag}{A^{\dagger}}
\newcommand{\Ahatdag}{\hat{A}^{\dagger}}
\newcommand{\Atdag}{\hat{A}^{(\tau)\dagger}}
\newcommand{\Bdag}{\hat{B}^{\dagger}}
\newcommand{\Bhat}{\hat{B}}
\newcommand{\Bhatdag}{\hat{B}^{\dagger}}
\newcommand{\Btdag}{\hat{B}^{(\tau)\dagger}}
\newcommand{\bhat}{\hat{b}}
\newcommand{\bdag}{b^{\dagger}}
\newcommand{\cdag}{c^{\dagger}}
\newcommand{\chat}{\hat{c}}
\newcommand{\chatdag}{\hat{c}^{\dagger}}
\newcommand{\degree}{^{\circ}}
\newcommand{\sprime}{s^{\prime}}
\newcommand{\Hhat}{\hat{H}}
\newcommand{\Hhatp}{\hat{H}^{\prime}}
\newcommand{\Ihat}{\hat{I}}
\newcommand{\Jhat}{\hat{J}}
\newcommand{\hhat}{\hat{h}}
\newcommand{\fp}{f^{(+)}}
\newcommand{\fpp}{f^{(+)\prime}}
\newcommand{\fm}{f^{(-)}}
\newcommand{\Fhat}{\hat{F}}
\newcommand{\Fhatdag}{\hat{F}^\dagger}
\newcommand{\Fhatp}{\hat{F}^{(+)}}
\newcommand{\Fhatm}{\hat{F}^{(-)}}
\newcommand{\Fhatpm}{\hat{F}^{(\pm)}}
\newcommand{\Fhatdagpm}{\hat{F}^{\dagger(\pm)}}
\newcommand{\Hc}{{\cal H}}
\newcommand{\Hcp}{{\cal H}^{\prime}}
\newcommand{\Ic}{{\cal I}}
\newcommand{\It}{\widetilde{I}}
\newcommand{\ITV}{{\cal I}_{\rm TV}}
\newcommand{\Jc}{{\cal J}}
\newcommand{\jp}{j^{\prime}}
\newcommand{\Qc}{{\cal Q}}
\newcommand{\Pc}{{\cal P}}
\newcommand{\Ec}{{\cal E}}
\newcommand{\Sc}{{\cal S}}
\newcommand{\Rc}{{\cal R}}

\newcommand{\ddg}{d^{\dagger}}

\newcommand{\Nhat}{\hat{N}}
\newcommand{\Nt}{\widetilde{N}}
\newcommand{\Vt}{\widetilde{V}}
\newcommand{\nL}[1]{n_{L_{#1}}}
\newcommand{\nK}[1]{n_{K_{#1}}}
\newcommand{\nKb}{\mbox{\boldmath $n_K$}}
\newcommand{\nLb}{\mbox{\boldmath $n_L$}}

\newcommand{\Res}{\text{Res}}

\newcommand{\mubar}{\bar{\mu}}

\newcommand{\Dc}{{\mathscr D}}
\newcommand{\Dp}{D^{(+)}}
\newcommand{\Ddag}{\hat{D}^{\dagger}}
\newcommand{\dhat}{\hat{d}}
\newcommand{\Dhat}{\hat{D}}
\newcommand{\Dhatp}{\hat{D}^{(+)}}
\newcommand{\Ghat}{\hat{G}}
\newcommand{\Glambda}{G^{(\lambda)}}
\newcommand{\Gstarlambda}{G^{(\lambda)\ast}}
\newcommand{\Qhat}{\hat{Q}}
\newcommand{\Rhat}{\hat{R}}
\newcommand{\Phat}{\hat{P}}
\newcommand{\Pdag}{\hat{P}^{\dagger}}
\newcommand{\Psihat}{\hat{\Psi}}
\newcommand{\Qdag}{Q^{\dagger}}
\newcommand{\That}{\hat{\Theta}}
\newcommand{\Thatt}{\widetilde{\hat{\Theta}}}
\newcommand{\Tr}{{\rm Tr}}

\newcommand{\ktilde}{\tilde{k}}

\newcommand{\Pcirc}{\stackrel{\circ}{P}}
\newcommand{\Qcirc}{\stackrel{\circ}{Q}}
\newcommand{\Ncirc}{\stackrel{\circ}{N}}
\newcommand{\Tcirc}{\stackrel{\circ}{\Theta}}
\newcommand{\Pcircp}{\stackrel{\circ}{P^{\prime}}}
\newcommand{\Qcircp}{\stackrel{\circ}{Q^{\prime}}}
\newcommand{\Ncircp}{\stackrel{\circ}{N^{\prime}}}
\newcommand{\Tcircp}{\stackrel{\circ}{\Theta^{\prime}}}
\newcommand{\Fp}{F^{(+)}}
\newcommand{\Fm}{F^{(-)}}
\newcommand{\Rp}{R^{(+)}}
\newcommand{\Rm}{R^{(-)}}
\newcommand{\Bt}{\widetilde{B}}
\newcommand{\lambdat}{\widetilde{\lambda}}
\newcommand{\Phatt}{\widetilde{\hat{P}}}
\newcommand{\ab}{\bf a}

\newcommand{\Ab}{{\mbox{\boldmath $\hat{A}$}}}
\newcommand{\Abdag}{{\mbox{\boldmath $\hat{A}$} }^{\dagger}}
\newcommand{\Bb}{{\mbox{\boldmath $\hat{B}$}}}
\newcommand{\cb}{\bf c}
\newcommand{\Db}{\mbox{\boldmath $D$}}
\newcommand{\Nb}{\mbox{\boldmath $N$}}
\newcommand{\Nbhat}{\hat{\mbox{\boldmath $N$}}}
\newcommand{\Qb}{\mbox{\boldmath $Q$}}
\newcommand{\Qhatt}{\widetilde{\hat{Q}}}
\newcommand{\Pb}{\mbox{\boldmath $P$}}
\newcommand{\phit}{\phi(t)}
\newcommand{\pdot}{\dot{p}}
\newcommand{\phix}[1]{\phi(#1)}
\newcommand{\qdot}{\dot{q}}
\newcommand{\phivib}{\phi(\eta^{\ast},\eta)}
\newcommand{\Ts}{{\cal T}}
\newcommand{\del}{\partial}
\newcommand{\eps}{\epsilon}
\newcommand{\beq}{\begin{equation}}
\newcommand{\beqa}{\begin{eqnarray}}
\newcommand{\eeq}{\end{equation}}
\newcommand{\eeqa}{\end{eqnarray}}
\newcommand{\Yb}{${}^{168}$Yb\ }
\newcommand{\Zhat}{\hat{Z}}
\newcommand{\rhodot}{\dot{\rho}}
\newcommand{\Khat}{\hat{K}}
\newcommand{\Kp}{K^{+}}
\newcommand{\Km}{K^{-}}
\newcommand{\Kz}{K^0}

\newcommand{\lb}{\bf l}
\newcommand{\sbold}{\bf s}

\newcommand{\Lp}{L^{+}}
\newcommand{\Lm}{L^{-}}
\newcommand{\Lz}{L^0}

\newcommand{\Mc}{{\cal M}}
\newcommand{\Mchat}{\hat{\cal M}}

\newcommand{\ddeta}{\frac{\partial}{\partial \eta}}
\newcommand{\ddetastar}{\frac{\partial}{\partial \eta^\ast}}
\newcommand{\etastar}{\eta^\ast}
\newcommand{\ketvib}{\ket{\phi (\etastar, \eta)}}
\newcommand{\bravib}{\bra{\phi (\etastar, \eta)}}
\newcommand{\zhateta}{\hat{z}(\eta)}
\newcommand{\zhat}{\hat{z}}
\newcommand{\oo}{\stackrel{\circ}{O}(\etastar,\eta)}
\newcommand{\oodag}{\stackrel{\circ}{O^{\dagger}}(\etastar,\eta)}
\newcommand{\oodagp}{\stackrel{\circ}{O^{\dagger\prime}}(\etastar,\eta)}
\newcommand{\oop}{\stackrel{\circ}{O^{\prime}}(\etastar,\eta)}
\newcommand{\Odag}{\hat{O}^{\dagger}}
\newcommand{\Ohat}{\hat{O}}
\newcommand{\Uinv}{U^{-1}(\etastar, \eta)}
\newcommand{\Uinvp}{U^{-1}(\etastar,\eta,\varphi,n)}
\newcommand{\U}{U(\etastar, \eta)}
\newcommand{\Up}{U(\etastar,\eta,\varphi,n)}
\newcommand{\etader}{\frac{\del}{\del \eta}}
\newcommand{\etastarder}{\frac{\del}{\del \etastar}}

\newcommand{\fb}{\mbox {\bfseries\itshape f}}
\newcommand{\SB}{\mbox {\bfseries\itshape S}}

\newcommand{\vbar}{\bar{v}}

\newcommand{\Udag}{U^{\dagger}}
\newcommand{\Vdag}{V^{\dagger}}

\newcommand{\Wc}{{\cal W}}
\newcommand{\Wcdag}{{\cal W}^{\dagger}}

\newcommand{\Xhat}{\hat{X}}
\newcommand{\Xdag}{\hat{X}^{\dagger}}

\renewcommand{\thanks}{\footnote}
\newcommand\tocite[1]{$^{\hbox{--}}$\cite{#1}}

\preprint{}

\title{Low energy collective modes of deformed superfluid nuclei within the  finite amplitude method
}

\author{Nobuo Hinohara}
\affiliation{%
Department of Physics and Astronomy, University of North Carolina, Chapel Hill, North Carolina, 27599-3255, USA
}
\affiliation{%
Department of Physics and Astronomy, University of Tennessee, Knoxville, Tennessee, 37996-1200, USA
}

\affiliation{%
Joint Institute for Heavy-Ion Research, Oak Ridge, Tennessee, 37831-6374, USA
}
\author{Markus Kortelainen}%
\affiliation{%
Department of Physics, P.O. Box 35 (YFL), FI-40014, University of Jyv\"{a}skyl\"{a}, Finland
}
\affiliation{%
Department of Physics and Astronomy, University of Tennessee, Knoxville, Tennessee, 37996-1200, USA
}
\author{Witold Nazarewicz}
\affiliation{%
Department of Physics and Astronomy, University of Tennessee, Knoxville, Tennessee, 37996-1200, USA
}
\affiliation{%
Physics Division, Oak Ridge National Laboratory, Oak Ridge, Tennessee 37831-6373, USA
}
\affiliation{%
Institute of Theoretical Physics, University of Warsaw, ul. Ho\.{z}a 69, PL-00-861, Warsaw, Poland
}

\date{\today}

\begin{abstract} 
\begin{description}
 \item[Background]
The major challenge for nuclear theory is to describe and predict global properties and collective modes of atomic nuclei.
 Of particular interest is the response of the nucleus to a
time-dependent external field that impacts the low-energy multipole and beta-decay strength.
\item[Purpose]
We propose a method to compute low-lying collective modes in deformed nuclei  within 
the finite amplitude method (FAM) based on the
quasiparticle random-phase approximation (QRPA). By using the analytic property 
of the response function, we find the QRPA  amplitudes by 
computing the residua of the FAM amplitudes by means of 
a contour integration around the QRPA poles in a complex frequency plane.
\item[Methods]
We use the superfluid nuclear density functional theory with  Skyrme energy density functionals, FAM-QRPA approach, and the conventional matrix formulation of the QRPA (MQRPA).
\item[Results]
We demonstrate  that the complex-energy FAM-QRPA  method reproduces low-lying collective states obtained within  the conventional 
matrix formulation of the QRPA theory. Illustrative calculations are performed for the isoscalar monopole strength in deformed $^{24}$Mg and for low-lying $K=0$ quadrupole vibrational modes of  deformed Yb and Er isotopes.
\item[Conclusions]
The proposed FAM-QRPA approach allows one to efficiently calculate  low-lying collective modes in spherical and deformed nuclei throughout the entire nuclear landscape, including shape-vibrational excitations, pairing vibrational modes, and beta-decay rates. 
\end{description}

\end{abstract}

\maketitle

%\tableofcontents

%%%%%%%%%%%%%%%%%%%%%%%%%%%%%%%%%%%%%%%%%%%%%%%%%%%%%%%%%%%%%%%%%%%%%%%%%%%%%%%%%%%%%%%%%%%%%%%%%%%%%%%%%%%%%%%%%%%%%%%%
%
%  Introduction
%
%%%%%%%%%%%%%%%%%%%%%%%%%%%%%%%%%%%%%%%%%%%%%%%%%%%%%%%%%%%%%%%%%%%%%%%%%%%%%%%%%%%%%%%%%%%%%%%%%%%%%%%%%%%%%%%%%%%%%%%%
\section{Introduction}

Vibrational modes of atomic nuclei provide crucial information about 
nuclear structure. In particular, collective low-lying states contain information about the nucleonic shell structure, pairing correlations, 
 and nuclear deformations \cite{Bohr-MottelsonV2,Ring-Schuck}. Giant resonances
tell us about global properties of nuclear matter, such as compressibility and symmetry energy \cite{Lip89,Har01}.
Electromagnetic strength plays an important role in nuclear reactions involving photo-nuclear
processes, including astrophysical reactions \cite{Arnould200797,PhysRevC.86.034328} and the transmutation of nuclear waste \cite{Bea12}.

The random-phase approximation (RPA) and its superfluid extension, the quasiparticle random-phase approximation (QRPA), are well-established microscopic theories 
describing excitations of many-body  systems \cite{Bohr-MottelsonV2,Ring-Schuck}. QRPA can be viewed  as a small-amplitude approximation of the time-dependent density functional theory \cite{Bla86,Nakatsukasa12}.
By using nuclear energy density functionals (EDFs) applicable to a large portion of the nuclear landscape, 
a variety of excited modes can be described by using QRPA.

Recently, there has been a considerable progress  in the area of fully self-consistent QRPA calculations based on the nuclear density functional theory.
Due to advances in high performance computing,  deformed QRPA frameworks have been developed that can address properties of well-bound and loosely-bound nuclei
\cite{PhysRevC.82.034324,PhysRevC.83.021304,PhysRevC.82.034326,PhysRevC.84.014332,PhysRevC.81.064307,PhysRevC.83.014314,peru:044313,PhysRevC.83.034309}.

The traditional QRPA methodology is based on a generalized eigenvalue problem  involving a QRPA matrix containing the residual two-quasiparticle interaction.
Because of a large number of quasiparticle states involved, the dimension of the QRPA matrix is typically quite 
large, especially  when spherical symmetry is broken. This makes the problem  computationally challenging.
Therefore, in order to reduce the dimension of the two-quasiparticle basis, additional cutoffs are imposed on the configuration space of 
 QRPA.  Such truncations result in   inconsistencies between the model spaces of Hartree-Fock-Bogoliubov (HFB) and  QRPA calculations, and can result in breaking self-consistency and appearance of spurious modes \cite{PhysRevC.71.034310}.

To circumvent these problems, efficient methods to solve RPA  have been formulated in the framework of the linear response 
theory and time-dependent HFB. One of these methods is the finite amplitude method (FAM) proposed in Ref.~\cite{nakatsukasa:024318}.
Within FAM, the strength function of an arbitrary one-body transition operator can be calculated without actually constructing and 
diagonalizing the full (Q)RPA matrix. Instead, the fields induced by the one-body transition (driving)  operator are calculated and the linear 
response problem is solved  iteratively. The practical implementation of the FAM  requires  minor extensions to the 
existing HFB codes to calculate the induced fields and, therefore, is fairly straightforward.
Systematic calculations with the FAM have been performed for the electric giant dipole resonances
and low-lying dipole strength, illustrating computational advantages of the method \cite{PhysRevC.80.044301,PhysRevC.84.021302}.
The FAM has also been extended to the superfluid systems, both  
spherical  \cite{PhysRevC.84.014314} and deformed \cite{PhysRevC.84.041305}.

Since the FAM equations are solved by introducing a small width, an imaginary part of the QRPA frequency,
the method is very effective for describing   excited modes in a region of high density of states. However, until now,
a direct application of  FAM to discrete low-lying excitations has not been fully accomplished.
Quite recently, an efficient  method to evaluate the QRPA matrix using FAM has been  reported \cite{PhysRevC.87.014331} that
 significantly reduces the computational effort, also
enabling computations of  low-lying discrete QRPA modes.
A disadvantage of this approach is that
a large memory is required to store the huge QRPA matrix, which subsequently needs to be diagonalized.

An alternative technique to solve the linear response  problem is based on the iterative Arnoldi diagonalization method
\cite{PhysRevC.81.034312}. This method was first implemented for spherical systems without pairing
and then further extended to spherical  superfluid nuclei \cite{PhysRevC.86.024303}.
Because the  Arnoldi diagonalization algorithm solves the QRPA eigenvalue problem in a smaller Krylov-space, the discrete excitations are within the scope of this method~\cite{PhysRevC.86.014307}.

The goal of this work is to derive a method to calculate the discrete low-lying QRPA modes within the FAM framework. We shall refer to this new technique as FAM-QRPA in the following.
Starting from the linear response theory, we show in Secs.~\ref{sec:fam} and \ref{sec:famdiscrete} that a contour integration in the complex frequency  plane around a QRPA root provides 
the QRPA eigenvectors. 
A similar technique was  proposed  in Ref.~\cite{Sakurai2003119} to solve generalized eigenvalue problems.
We devise several techniques to  compute and assess
the accuracy of QRPA modes.
Next, in  Sec.~\ref{sec:result},  we numerically demonstrate that the discrete FAM-QRPA solution for the low-lying states 
reproduces the modes  obtained within the conventional matrix formulation of QRPA (MQRPA) and we apply FAM-QRPA
to collective modes in  deformed Er and Yb nuclei.
Finally, the conclusions of our work are given in Sec.~\ref{sec:conclusion}.

%%%%%%%%%%%%%%%%%%%%%%%%%%%%%%%%%%%%%%%%%%%%%%%%%%%%%%%%%%%%%%%%%%%%%%%%%%%%%%%%%%%%%%%%%%%%%%%%%%%%%%%%%%%%%%%%%%%%%%%%
%
%  Finite amplitude method
%
%%%%%%%%%%%%%%%%%%%%%%%%%%%%%%%%%%%%%%%%%%%%%%%%%%%%%%%%%%%%%%%%%%%%%%%%%%%%%%%%%%%%%%%%%%%%%%%%%%%%%%%%%%%%%%%%%%%%%%%%
\section{Finite amplitude method} \label{sec:fam}

In this section we recapitulate the derivation of the FAM equations 
for superfluid systems following Sec.~II of Ref.~\cite{PhysRevC.84.014314}.
In the FAM formalism, the polarization of the system is induced by an external time-dependent field $\Fhat(t)$ with a frequency $\omega$: 
%%%
\begin{equation}\label{eq:F}
  \Fhat(t)  =  \eta \left\{ \Fhat e^{-i\omega t} + \Fhat^\dagger  e^{i\omega t} \right\},
\end{equation}  
where  
\begin{equation}\label{eq:Fhatb}    
  \Fhat      =  \frac{1}{2} \sum_{\mu\nu} 
\left\{ F^{20}_{\mu\nu} \Abdag_{\mu\nu}  + F^{02}_{\mu\nu} \Ab_{\mu\nu} +  F^{11}_{\mu\nu} \Bb_{\mu\nu} \right\},
\end{equation}
and  $\Abdag_{\mu\nu}=\ahatdag_\mu \ahatdag_\nu$ and $\Bb_{\mu\nu}=\ahatdag_\mu \ahat_\nu$ are two-quasi\-par\-ti\-cle operators.
The parameter $\eta$ is a (small) real number to expand particle and pair HFB densities to the first order. Contrary to Ref.~\cite{PhysRevC.84.014314},
we assume that  $\Fhat$ is $\omega$-independent in all applications in this work.
However, our scheme can be easily extended to the case where $\Fhat$  depends on  $\omega$.

The time-evolution of  quasiparticle operators  under the external field $\Fhat(t)$ is determined by the time-dependent HFB (TDHFB) equation:
\begin{align}
  i \frac{\del}{\del t}\ahat_{\mu}(t) = [ \Hhat(t) + \Fhat(t), \ahat_{\mu}(t) ] \, , \label{eq:TDHFB}
\end{align}
where time-dependent oscillation of quasiparticle operators is:
\begin{subequations}\begin{align}
  \ahat_{\mu}(t) & = \left\{ \ahat_\mu + \delta \ahat_\mu(t)\right\} e^{iE_\mu t} \, ,\\
 \delta \ahat_\mu(t) & = \eta \sum_{\nu} \ahatdag_\nu \left\{ X_{\nu\mu}(\omega) e^{-i\omega t} + Y^\ast_{\nu\mu}(\omega) e^{i\omega t}\right\} \, ,
\end{align}\label{eq:a}\end{subequations}
where $E_\mu$ is the one-quasiparticle energy and $X_{\mu\nu}(\omega)$ and $Y_{\mu\nu}(\omega)$ are the FAM amplitudes.

In terms of time-dependent quasiparticles, the TDHFB Hamiltonian can be written as $\Hhat(t)= \Hhat_0 + \delta \Hhat(t)$,
where: 
\begin{equation}
  \Hhat_0 = \sum_{\mu} E_\mu \Bb_{\mu\mu}
\end{equation}
is the HFB Hamiltonian and 
\begin{equation}\label{eq:H}
  \delta \Hhat (t)  = \eta \left\{ \delta \Hhat(\omega) e^{-i\omega t} + \delta \Hhat^\dagger (\omega) e^{i\omega t}\right\} 
\end{equation}
with
\begin{equation}
  \delta \Hhat (\omega) = \frac{1}{2} \sum_{\mu\nu} \left\{
  \delta H^{20}_{\mu\nu}(\omega) \Abdag_{\mu\nu} + \delta H^{02}_{\mu\nu}(\omega) \Ab_{\mu\nu} \right\}
\end{equation}
represents a  small-amplitude oscillation.

Inserting (\ref{eq:F}), (\ref{eq:a}), and (\ref{eq:H}) into (\ref{eq:TDHFB}) results in the FAM equations:
\begin{subequations}\begin{align}
  (E_{\mu} + E_{\nu} - \omega) X_{\mu\nu}(\omega)  + \delta H^{20}_{\mu\nu}(\omega) & = -F^{20}_{\mu\nu} \, ,\\
  (E_{\mu} + E_{\nu} + \omega) Y_{\mu\nu}(\omega)  + \delta H^{02}_{\mu\nu}(\omega) & = -F^{02}_{\mu\nu} \, .
\end{align}\label{eq:FAM}\end{subequations} 
By expanding $\delta H^{20}(\omega)$ and $\delta H^{02}(\omega)$ in terms of $X(\omega)$ and $Y(\omega)$, one obtains:
\begin{subequations}\begin{align}
\delta H^{20}_{\mu\nu}(\omega) & = \sum_{\mu'<\nu'} 
\left\{ A_{\mu\nu,\mu'\nu'} - (E_\mu + E_\nu) \delta_{\mu\mu'} \delta_{\nu\nu'} \right\} X_{\mu'\nu'}(\omega) \nonumber \\
&+ \sum_{\mu'<\nu'} B_{\mu\nu,\mu'\nu'} Y_{\mu'\nu'}(\omega) \, , \\
\delta H^{02}_{\mu\nu}(\omega) &=  
\sum_{\mu'<\nu'}\left\{ A^\ast_{\mu\nu,\mu'\nu'} - (E_\mu + E_\nu) \delta_{\mu\mu'}\delta_{\nu\nu'}
\right\} Y_{\mu'\nu'}(\omega) \nonumber \\
&+ \sum_{\mu'<\nu'} B^{\ast}_{\mu\nu,\mu'\nu'} X_{\mu'\nu'}(\omega) \, ,
\end{align}\label{eq:dHAB}\end{subequations}
where $A$ and $B$ are the usual QRPA matrices \cite{Ring-Schuck}. 
The advantage of the FAM formulation is that the $A$ and $B$ matrices do not have to be computed explicitly.
By substituting (\ref{eq:dHAB}) into the FAM equations (\ref{eq:FAM}),
the linear response equation becomes:
\begin{align}
  \left[
    \begin{pmatrix} A & B \\ B^\ast & A^\ast \end{pmatrix}
    - \omega
    \begin{pmatrix} 1 & 0 \\ 0 & -1 \end{pmatrix}
    \right] 
  \begin{pmatrix} X(\omega) \\ Y(\omega) \end{pmatrix}
  = -
  \begin{pmatrix} F^{20}  \\ F^{02}  \end{pmatrix}, \label{eq:linres}
\end{align}
where the sum over two quasiparticle space is restricted to quasiparticle indices $\mu<\nu$. The FAM equations are thus equivalent to the linear response formalism. Furthermore,
the left hand side of (\ref{eq:linres}) yields the QRPA equations when the right-hand side is set to zero.
The FAM equations (\ref{eq:FAM}) are solved by using complex frequencies $\omega_\gamma=\omega+i\gamma$, where
the imaginary part $\gamma$ corresponds to a smearing width.

In terms of  the FAM amplitudes $X(\omega_\gamma)$ and $Y(\omega_\gamma)$,
the strength function $dB(\omega ;F)/d\omega$ for the operator $\Fhat$ can be written as:
\begin{align}
  \frac{dB(\omega;F)}{d\omega} & = -\frac{1}{\pi}{\rm Im} S(F;\omega_\gamma), \\
  S(F;\omega_\gamma) & = \sum_{\mu<\nu} \left\{
  F^{20\ast}_{\mu\nu} X_{\mu\nu}(\omega_\gamma) + F^{02\ast}_{\mu\nu} Y_{\mu\nu}(\omega_\gamma)
\right\}. \label{eq:FAMstrength}
\end{align}

%%%%%%%%%%%%%%%%%%%%%%%%%%%%%%%%%%%%%%%%%%%%%%%%%%%%%%%%%%%%%%%%%%%%%%%%%%%%%%%%%%%%%%%%%%%%%%%%%%%%%%%%%%%%%%%%%%%%%%%%
%
%  FAM for discrete QRPA modes
%
%%%%%%%%%%%%%%%%%%%%%%%%%%%%%%%%%%%%%%%%%%%%%%%%%%%%%%%%%%%%%%%%%%%%%%%%%%%%%%%%%%%%%%%%%%%%%%%%%%%%%%%%%%%%%%%%%%%%%%%%
\section{FAM for discrete QRPA modes} \label{sec:famdiscrete}

The objective of this work is to formulate a FAM capable of describing low-lying discrete QRPA modes.
We start by introducing the  $2N\times 2N$ matrices \cite{Ring-Schuck}:
\begin{align}\label{matrices}
  {\cal S}=\begin{pmatrix} A & B \\ B^\ast & A^\ast \end{pmatrix}, \,
  {\cal N}=\begin{pmatrix} 1 & 0 \\ 0 & -1 \end{pmatrix}, \,
  {\cal X}=\begin{pmatrix} X & Y^\ast \\ Y & X^\ast \end{pmatrix},
\end{align}
where $N$ is the dimension of the two-quasiparticle space, and the matrix elements
$X_{\mu\nu}^i$ and $Y_{\mu\nu}^i$ of ${\cal X}$ are the QRPA amplitudes of the $i$-th mode with a positive eigenfrequency $\Omega_i$.
There also exists a counterpart QRPA solution ($Y^{i\ast}, X^{i\ast}$) with a negative eigenfrequency $-\Omega_i$.
We assume here that all the QRPA frequencies are real, that is, ${\cal S}$ is positive definite.
In terms of matrices (\ref{matrices}), the QRPA equation can be expressed as:
\begin{align}
  {\cal S}{\cal X} = {\cal N}{\cal X}{\cal O}, \label{eq:QRPA}
\end{align}
where ${\cal O}$ is a $2N\times 2N$ diagonal matrix containing the QRPA eigenfrequencies
($\Omega_1, \ldots, \Omega_N, -\Omega_1, \ldots, -\Omega_N$).
The orthonormalization condition for the QRPA eigenvectors is:
\begin{align}
  {\cal X}{\cal N}{\cal X}^\dagger = {\cal N}. \label{eq:orthonormalization}
\end{align}
By applying Eqs.~(\ref{eq:QRPA}) and (\ref{eq:orthonormalization}), the matrix on the left-hand side of (\ref{eq:linres})
can be inverted, which yields the FAM amplitudes \cite{Ring-Schuck}:
\begin{align}
\begin{pmatrix}X(\omega_\gamma) \\ Y(\omega_\gamma) \end{pmatrix}
& = -R(\omega_\gamma) \begin{pmatrix} F^{20} \\ F^{02} \end{pmatrix} \nonumber \\
& = -{\cal X}({\cal O} - \omega_\gamma {\cal I})^{-1} {\cal N}{\cal X}^\dagger \begin{pmatrix} F^{20} \\ F^{02} \end{pmatrix},\label{eq:FAMXY1}
\end{align}
where ${\cal I}$ is a $2N\times 2N$ unit matrix and $R(\omega_\gamma)$ is the response function. 
The explicit form of $R(\omega_\gamma)$ is:
\begin{widetext}
\begin{equation}
  R_{\mu\nu\mu'\nu'}(\omega_\gamma)
  = \sum_i
\begin{bmatrix}
  \displaystyle\frac{X_{\mu\nu}^i X_{\mu'\nu'}^{i\ast}}{\Omega_i - \omega_\gamma} + \frac{Y_{\mu\nu}^{i\ast} Y_{\mu'\nu'}^{i}}{\Omega_i + \omega_\gamma}
  &
  \displaystyle\frac{X_{\mu\nu}^i Y_{\mu'\nu'}^{i\ast}}{\Omega_i - \omega_\gamma} + \frac{Y_{\mu\nu}^{i\ast} X_{\mu'\nu'}^{i}}{\Omega_i + \omega_\gamma} 
  \\
  \displaystyle\frac{Y_{\mu\nu}^i X_{\mu'\nu'}^{i\ast}}{\Omega_i - \omega_\gamma} + \frac{X_{\mu\nu}^{i\ast} Y_{\mu'\nu'}^{i}}{\Omega_i + \omega_\gamma}
  &
  \displaystyle\frac{Y_{\mu\nu}^i Y_{\mu'\nu'}^{i\ast}}{\Omega_i - \omega_\gamma} + \frac{X_{\mu\nu}^{i\ast} X_{\mu'\nu'}^{i}}{\Omega_i + \omega_\gamma}
\end{bmatrix}. \label{eq:resp}
\end{equation}
\end{widetext}
Substitution of Eq.~(\ref{eq:resp}) into Eq.~(\ref{eq:FAMXY1}) provides the relation between the FAM amplitudes and QRPA amplitudes
\begin{subequations}
\begin{align}
  X_{\mu\nu}(\omega_\gamma) & = -\sum_i \left\{
  \frac{ X^i_{\mu\nu} \bra{i}\Fhat\ket{0}} {\Omega_i - \omega_\gamma} + \frac{ Y^{i\ast}_{\mu\nu} \bra{0}\Fhat\ket{i}}{\Omega_i + \omega_\gamma} 
 \right\}, \\
  Y_{\mu\nu}(\omega_\gamma) & = -\sum_i \left\{
  \frac{ Y^i_{\mu\nu} \bra{i}\Fhat\ket{0}} {\Omega_i - \omega_\gamma} + \frac{ X^{i\ast}_{\mu\nu} \bra{0}\Fhat\ket{i}}{\Omega_i + \omega_\gamma}
 \right\},
\end{align}
\label{eq:FAMXY2}
\end{subequations}
where
\begin{align}
  \bra{i}\Fhat\ket{0} & = \bra{\Phi_0}[\Ohat_i,\Fhat]\ket{\Phi_0} \nonumber \\
   & = \sum_{\mu<\nu}(X^{i\ast}_{\mu\nu} F^{20}_{\mu\nu} + Y^{i\ast}_{\mu\nu} F^{02}_{\mu\nu})\, ,\label{eq:strengthfromQRPA} \\
  \bra{0}\Fhat\ket{i} & = \bra{\Phi_0}[\Ohat^\dagger_i,\Fhat]\ket{\Phi_0} \nonumber \\
   & = \sum_{\mu<\nu}(F^{02}_{\mu\nu} X^{i}_{\mu\nu} + F^{20}_{\mu\nu} Y^{i}_{\mu\nu}),
\end{align}
are the QRPA transition strengths between the QRPA ground state $\ket{0}$ and $i$-th excited state $\ket{i}$, 
 $\ket{\Phi_0}$ is the HFB state, the operator:
\begin{align}
\Ohat^\dagger_i = \sum_{\mu<\nu} \{ X^i_{\mu\nu} \Abdag_{\mu\nu} - Y^i_{\mu\nu} \Ab_{\mu\nu} \}
\end{align} 
is the QRPA phonon operator, and $X^i_{\mu\nu}$ and $Y^i_{\mu\nu}$ are the QRPA amplitudes of a  state $i$.

Equation (\ref{eq:FAMXY2})  shows that the FAM amplitudes $X(\omega_\gamma)$ and $Y(\omega_\gamma)$ have 
first-order poles on the real axis at $\omega_\gamma=\Omega_i$ and $-\Omega_i$. By calculating 
the standard FAM strength function, approximate positions of the poles of the low-lying states of interest
can be located. This allows one to define a closed contour $C_i$ in the complex energy plane that encloses the $i$-th positive pole $\Omega_i$.
According to Cauchy's integral formula, the contour integration of the FAM amplitudes (\ref{eq:FAMXY2})  along $C_i$  gives the residue at  the $i$-th pole:
\begin{subequations}\begin{align}
  \frac{1}{2\pi i} \oint_{C_i} X_{\mu\nu}(\omega_\gamma) d\omega_\gamma & =  
  \Res(X_{\mu\nu},\Omega_i)=
  X^i_{\mu\nu}\bra{i}\Fhat\ket{0}, \\
  \frac{1}{2\pi i} \oint_{C_i} Y_{\mu\nu}(\omega_\gamma) d\omega_\gamma & =  
   \Res(Y_{\mu\nu},\Omega_i) = Y^i_{\mu\nu}\bra{i}\Fhat\ket{0}.
\end{align}\label{eq:FAMXYcontour} 
\end{subequations}
The absolute value of the transition strength for the $i$-th QRPA mode can then be expressed  as:
\begin{align}
 |\bra{i}\Fhat\ket{0}|^2 & = \sum_{\mu<\nu} \left\{
 \Big| \frac{1}{2\pi i} \oint_{C_i} X_{\mu\nu}(\omega_\gamma)d\omega_\gamma\Big|^2 \right. \nonumber \\
&- \left.
 \Big| \frac{1}{2\pi i} \oint_{C_i} Y_{\mu\nu}(\omega_\gamma)d\omega_\gamma\Big|^2
\right\}, \label{eq:strengthfromXY}
\end{align}
where we have used the normalization condition (\ref{eq:orthonormalization}) for the QRPA amplitudes. The individual QRPA amplitudes 
$X_{\mu\nu}^i$ and $Y_{\mu\nu}^i$ can thus be calculated as:
\begin{subequations}\begin{align}
  X_{\mu\nu}^i & = e^{-i\theta} |\bra{i}\Fhat\ket{0}|^{-1} \frac{1}{2\pi i}\oint_{C_i} X_{\mu\nu}(\omega_\gamma)d\omega_\gamma, \\
  Y_{\mu\nu}^i & = e^{-i\theta} |\bra{i}\Fhat\ket{0}|^{-1} \frac{1}{2\pi i}\oint_{C_i} Y_{\mu\nu}(\omega_\gamma)d\omega_\gamma.
\end{align}\label{eq:QRPAdiscreteXY}\end{subequations}
The common phase $e^{i\theta}=\bra{i}\Fhat\ket{0}/\vert\bra{i}\Fhat\ket{0}\vert$
cannot be determined and remains arbitrary.

The information about the exact value of the QRPA eigenfrequency is not necessary to perform the contour integration as long 
as the corresponding pole  is located inside the contour. However, it can be calculated from the integration of the induced fields. Indeed, from 
Eqs.~(\ref{eq:dHAB}) and (\ref{eq:QRPA}), one obtains:
\begin{subequations}\begin{align}
  \frac{1}{2\pi i} \oint_{C_i} \delta H^{20}_{\mu\nu}(\omega_\gamma) d\omega_\gamma & = 
 \bra{i}\Fhat\ket{0} X^i_{\mu\nu} \left\{ \Omega_i - (E_\mu + E_\nu)\right\}, \\
  \frac{1}{2\pi i} \oint_{C_i} \delta H^{02}_{\mu\nu}(\omega_\gamma) d\omega_\gamma & = 
 \bra{i}\Fhat\ket{0} Y^i_{\mu\nu} \left\{-\Omega_i - (E_\mu + E_\nu)\right\}.
\end{align} \label{eq:omega1}\end{subequations}
These $2N$ equations can be used to compute $\Omega_i$,
but this method is prone to large numerical errors when amplitudes $X^i_{\mu\nu}$ or $Y^i_{\mu\nu}$ are very small. To this end, a better way
of determining the QRPA eigenfrequencies is
through an expression derived from Eq.~(\ref{eq:omega1}):
\begin{align}\label{eq:omega2}
  \Omega_i^2 & = \sum_{\mu<\nu} ( |\Omega_i X^i_{\mu\nu}|^2 - |\Omega_i Y^i_{\mu\nu}|^2) = \frac{1}{|\bra{i}\Fhat\ket{0}|^2} \sum_{\mu<\nu} \nonumber \\
& \left\{ 
\Big| \frac{1}{2\pi i} \oint_{C_i} \left((E_\mu + E_\nu) X_{\mu\nu}(\omega_\gamma) + \delta H^{20}_{\mu\nu}(\omega_\gamma)\right)d\omega_\gamma \Big|^2\right. \nonumber \\
-& \left. 
\Big| \frac{1}{2\pi i} \oint_{C_i} \left((E_\mu + E_\nu) Y_{\mu\nu}(\omega_\gamma) + \delta H^{02}_{\mu\nu}(\omega_\gamma)\right)d\omega_\gamma \Big|^2\right\}.
\end{align}

The formalism presented above allows one to establish an explicit connection between
the FAM strength function and the smeared QRPA strength function. By
substituting Eq.~(\ref{eq:FAMXY2}) into Eq.~(\ref{eq:FAMstrength}) we obtain:
\begin{align}
  S(F,\omega_\gamma) & = -\sum_i \left( \frac{|\bra{i}\Fhat\ket{0}|^2}{\Omega_i - \omega - i\gamma}
 + \frac{|\bra{i}\Fhat\ket{0}|^2}{\Omega_i + \omega + i\gamma} \right), \label{eq:FAMstrength2} \\
 \frac{dB}{d\omega}(F,\omega) & = -\frac{1}{\pi} {\rm Im} S(F,\omega_\gamma) \nonumber \\
& = \frac{\gamma}{\pi} 
\sum_i \left\{
\frac{|\bra{i}\Fhat\ket{0}|^2}{(\Omega_i -\omega)^2 + \gamma^2}
 - \frac{|\bra{i}\Fhat\ket{0}|^2}{(\Omega_i + \omega)^2 + \gamma^2} \right\}.
\end{align}
According to Eq.~(\ref{eq:FAMstrength2}), the discrete QRPA transition strength can be directly computed
from the FAM strength function (\ref{eq:FAMstrength}):
\begin{align}
|\bra{i}\Fhat\ket{0}|^2  = \frac{1}{2\pi i} \oint_{C_i} S(F,\omega)d\omega \, . \label{eq:strengthfromFAMS}
\end{align}

In summary, as discussed above, there  exist several techniques, 
based on the residue at the QRPA pole, to calculate discrete transition strengths within the FAM-QRPA formalism:
\begin{enumerate}[A:]
\item The contour integration of the FAM amplitudes $X_{\mu\nu}(\omega)$ and $Y_{\mu\nu}(\omega)$ as in Eq.~(\ref{eq:strengthfromXY});
\item The contour integration of the FAM strength function as in Eq.~(\ref{eq:strengthfromFAMS});
\item Individual QRPA amplitudes $X^{i}_{\mu\nu}$ and $Y^{i}_{\mu\nu}$ can be found using (\ref{eq:QRPAdiscreteXY})
      to obtain the transition matrix element (\ref{eq:strengthfromQRPA});  
\item The  QRPA amplitudes $X^{i}_{\mu\nu}$ and $Y^{i}_{\mu\nu}$ found with technique C are independent of the choice of the 
      external field used in FAM-QRPA. Therefore, for example, the isoscalar  strength associated with the field $\Fhat'$ can be computed using the 
      QRPA amplitudes obtained in FAM-QRPA  with the isovector  external field $\Fhat$.
\end{enumerate}
Although all of these strategies are formally equivalent, the technique B is  the easiest to implement in the current FAM codes. By  virtue of D, once the discrete QRPA amplitudes have been found for a given state, they can be  used to calculate a transition matrix element for any transition  operator.

If assigned incorrectly, the integration contour  $C'$  could  include  secondary unwanted  poles. (For example, there could 
be two states: a collective one carrying  a strong transition strength and a  nearby-lying  non-collective one with a negligible contribution 
to the total transition strength.) 
Since the FAM amplitudes $X_{\mu\nu}(\omega)$ and $Y_{\mu\nu}(\omega)$ (\ref{eq:FAMXY2}) are sums of  the residua,
the right hand side of Eq.~(\ref{eq:FAMXYcontour}) contains contributions from all the poles included inside $C'$.
The calculated transition strength then becomes:
\begin{align}
B(C';F) = \sum_{i\in C'}|\bra{i}\Fhat\ket{0}|^2.
\end{align}
Because of the orthogonality of QRPA amplitudes $X_{\mu\nu}^{i}$ and $Y_{\mu\nu}^{i}$, the interference terms between 
different states cancel out. Therefore,
if  $C'$ encircles two or more poles, the transition strengths from all those  poles contribute to the total strength without the interference term when techniques   A and B are used. Within C, calculated discrete amplitudes 
$X^{i}_{\mu\nu}$ and $Y^{i}_{\mu\nu}$  contain a mixture of all states inside the contour. However, when applied to Eq.~(\ref{eq:strengthfromQRPA}),  the same transition strength as with techniques  A and B is obtained.
However, in the method D, due to the incorrect amplitudes $X^{i}_{\mu\nu}$ and $Y^{i}_{\mu\nu}$, the final strength
\begin{align}
B_{\rm D}(C';F') = \frac{ \vert\sum_{i\in C'}\bra{i}\Fhat'\ket{0}\bra{i}\Fhat\ket{0}^*\vert^2 }
{ \sum_{i\in C'} \vert\bra{i}\Fhat\ket{0}\vert^2 },
\end{align}
would be incorrect.

As will be demonstrated in Sec.~\ref{sec:result}, we have checked numerically that 
when the contour includes multiple poles, techniques A-C
indeed yield the total summed strength while D  does not. This apparent deficiency of D can be used to our advantage to verify that the selected contour $C_{i}$ includes only one pole.

One can also find a clue to correct the assignment of the contour by calculating the QRPA eigenfrequency $\Omega^2$ using Eq.~(\ref{eq:omega2}):
\begin{equation}
\Omega_{C'}^{2} = \sum_{i\in C'} \left(\vert\bra{i}\Fhat\ket{0}\vert^2\Omega_i^2\right) \big/ \sum_{i\in C'}\vert\bra{i}\Fhat\ket{0}\vert^2. 
\end{equation}
When the contour encloses one collective and one non-collective QRPA root with respect to an external field $\Fhat$, 
Eq.~(\ref{eq:omega2}) yields the approximate energy of the collective state.

%%%%%%%%%%%%%%%%%%%%%%%%%%%%%%%%%%%%%%%%%%%%%%%%%%%%%%%%%%%%%%%%%%%%%%%%%%%%%%%%%%%%%%%%%%%%%%%%%%%%%%%%%%%%%%%%%%%%%%%%
%
%  Results
%
%%%%%%%%%%%%%%%%%%%%%%%%%%%%%%%%%%%%%%%%%%%%%%%%%%%%%%%%%%%%%%%%%%%%%%%%%%%%%%%%%%%%%%%%%%%%%%%%%%%%%%%%%%%%%%%%%%%%%%%%
\section{Numerical results} \label{sec:result}

To validate the  FAM-QRPA formalism discussed in the previous section,
we carried out numerical computations  using the FAM framework developed in Ref.~\cite{PhysRevC.84.041305} to evaluate
the  transition strength and the corresponding residua.
Our FAM-QRPA  method is based on the HFB code \pr{HFBTHO} \cite{Stoitsov200543,Stoitsov20131592}, which solves
the Skyrme-HFB equations  in the (transformed) harmonic oscillator basis assuming  axial and mirror symmetries.
The FAM equations are solved iteratively by using the modified Broyden's procedure \cite{PhysRevB.38.12807,baran:014318}, 
which offers a rapid and stable convergence, which weakly depends on the magnitude of the imaginary frequency $\gamma$.

%%%%%%%%%%%%%%%%%%%%%%%%%%%%%%%%%%%%%%%%%%%%%%%%%%%%%%%%%%%%%%%%%%%%%%%%%%%%%%%%%%%%%%%%%%%%%%%%%%%%%%%%%%%%%%%%%%%%%%%%
%
%  Numerical calculations, 24Mg
%
%%%%%%%%%%%%%%%%%%%%%%%%%%%%%%%%%%%%%%%%%%%%%%%%%%%%%%%%%%%%%%%%%%%%%%%%%%%%%%%%%%%%%%%%%%%%%%%%%%%%%%%%%%%%%%%%%%%%%%%%
\subsection{Test case: monopole strength in $^{24}$Mg}

To compare with full  MQRPA, we consider the same  case of the monopole strength in $^{24}$Mg as discussed in Ref.~\cite{PhysRevC.84.041305}.
We use SLy4 Skyrme EDF \cite{Chabanat1998231}  and a contact volume pairing
with a 60\,MeV quasiparticle energy cutoff and the  pairing strength  $V_0=-125.20\,{\rm MeV\, fm}^{-3}$  for both neutrons and protons. In order to perform exact comparison without any truncation at the MQRPA level, we take
the single-particle basis consisting of $N_{\rm sh}=5$ oscillator shells
~\cite{PhysRevC.84.041305}.

The oblate-deformed HFB minimum of $^{24}$Mg was obtained at the
quadrupole mass deformation $\beta=-0.163$. In this configuration, both neutrons and protons  are in the superfluid phase, 
with pairing gaps $\Delta_{\rm n}=0.666$\,MeV and $\Delta_{\rm p}=0.654$\,MeV, respectively.
In the FAM calculation, we used the value of the parameter $\eta=10^{-7}$, which was found to provide the best accuracy~\cite{PhysRevC.84.041305}. For the convergence criterion of  FAM iterations,  defined in terms of the maximum difference between collective FAM amplitudes in two consecutive iterations,
we used the value of $10^{-5}$; this accuracy 
is typically reached after about 40 iterations.

As for $\Fhat$, we consider the isoscalar monopole (ISM) and isovector monopole (IVM) operators:
\begin{subequations}
\begin{align}
  \Fhat^{\rm ISM} & =  \frac{eZ}{A}\sum_{i=1}^A r_i^2, \\
  \Fhat^{\rm IVM} &  =  \frac{eZ}{A}\sum_{i=1}^N r_i^2 - \frac{eN}{A}\sum_{i=1}^Z r_i^2.
\end{align}
\end{subequations}
For the integration contours we take  circles with radii 0.02\,MeV, centered close to  
MQRPA frequencies. The contour integration is discretized with 11 points, unless
stated otherwise.
%%%%%%%%%%%%%%%%%%%%%%%%%%%%%%%%%%%%%%%%%%%%%%%%%%%%%%%%%%%%%%%%%%%%%
% Figure 1 begin
%%%%%%%%%%%%%%%%%%%%%%%%%%%%%%%%%%%%%%%%%%%%%%%%%%%%%%%%%%%%%%%%%%%%%
\begin{figure}[htbp]
\includegraphics[width=1.0\columnwidth]{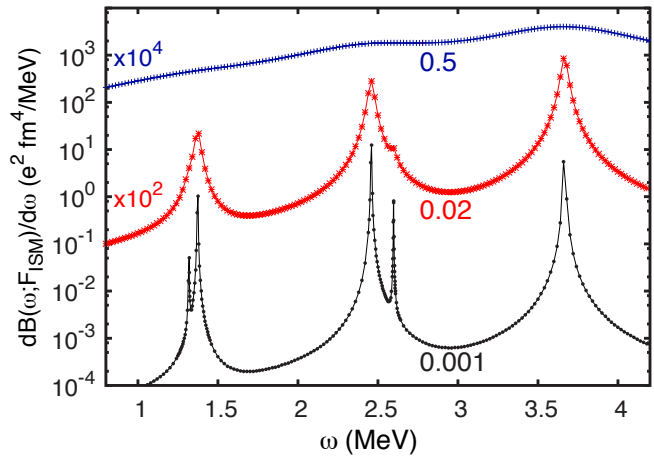}
\caption{(Color online)
The low-lying isoscalar monopole strength at the oblate HFB minimum of $^{24}$Mg 
calculated with the conventional FAM-QRPA using three 
 values of smearing width $\gamma$ (in MeV).
\label{fig:24Mg-ISM}}
\end{figure}
%%%%%%%%%%%%%%%%%%%%%%%%%%%%%%%%%%%%%%%%%%%%%%%%%%%%%%%%%%%%%%%%%%%%%
% Figure 1 end
%%%%%%%%%%%%%%%%%%%%%%%%%%%%%%%%%%%%%%%%%%%%%%%%%%%%%%%%%%%%%%%%%%%%%

Figure~\ref{fig:24Mg-ISM} shows the isoscalar monopole strength function at the oblate configuration  of $^{24}$Mg calculated 
with the conventional FAM by using  three values of $\gamma$.
The strength function obtained with $\gamma=0.5$\,MeV shows a very smooth distribution with the broad bumps carrying the largest strength. By going to
smaller values of $\gamma$, one  reveals the detailed structure of QRPA modes. For example, to 
separate the smaller first peak at $\Omega_1=1.32$\,MeV from the second one at $\Omega_2=1.37$\,MeV, a 
very small $\gamma$ -- of the order of 1\,keV -- is required.

%%%%%%%%%%%%%%%%%%%%%%%%%%%%%%%%%%%%%%%%%%%%%%%%%%%%%%%%%%%%%%%%%%%%%
% Tables I and II begin
%%%%%%%%%%%%%%%%%%%%%%%%%%%%%%%%%%%%%%%%%%%%%%%%%%%%%%%%%%%%%%%%%%%%%
\begin{table*}[htbp]
\caption{\label{table:24MgIS}
Low-lying $K=0$ QRPA energies $\Omega_i$ and  isoscalar monopole strength $|\bra{i}\Fhat\ket{0}|^2$ 
calculated with  MQRPA and FAM-QRPA  for the oblate configuration of $^{24}$Mg. 
All the modes with $\Omega_i <  7.5$\,MeV are listed. The transition strength
was computed using the techniques A-D described in Sec.~\ref{sec:famdiscrete}.
The isoscalar monopole strength FAM-D is calculated from the FAM-QRPA amplitudes 
generated by the external isovector monopole  field.  The numbers in parentheses denote powers of 10.}
\begin{ruledtabular}
\begin{tabular}{cc|lllll}
\multicolumn{2}{c|}{$\Omega_i$ (MeV)}    & \multicolumn{5}{c}{$|\bra{i}\Fhat^{\rm ISM}\ket{0}|^2$ ($e^2\,{\rm fm}^4$)}       \\
MQRPA & FAM & MQRPA   & FAM-A     
                    & FAM-B   & FAM-C     
                    & FAM-D    \\ \hline
1.3185 & 1.3183 & 5.729(-4) & 5.771(-4) & 5.773(-4) & 5.776(-4) & 5.781(-4) \\
1.3731 & 1.3731 & 1.539(-2) & 1.511(-2) & 1.511(-2) & 1.510(-2) & 1.511(-2) \\
2.4582 & 2.4581 & 0.1796    & 0.1780    & 0.1782    & 0.1784    & 0.1783    \\
2.5998 & 2.5975 & 2.957(-3) & 3.056(-3) & 3.058(-3) & 3.060(-3) & 3.057(-3) \\
3.6687 & 3.6657 & 0.5776    & 0.5755    & 0.5771    & 0.5788    & 0.5788    \\
5.1185 & 5.1212 & 3.539(-4) & 3.744(-4) & 4.040(-4) & 4.360(-4) & 4.345(-4) \\ 
7.4108 & 7.4084 & 0.4900    & 0.4820    & 0.4834    & 0.4848    & 0.4848    \\
\end{tabular}
\end{ruledtabular}
\end{table*}

\begin{table*}[htbp]
\caption{\label{table:24MgIV}
Similar as in  Table~\ref{table:24MgIS} but for the isovector monopole modes.}
\begin{ruledtabular}
\begin{tabular}{cc|lllll}
\multicolumn{2}{c|}{$\Omega_i$ (MeV)}    & \multicolumn{5}{c}{$|\bra{i}\Fhat^{\rm IVM}\ket{0}|^2$ ($e^2\,{\rm fm}^4$)}       \\
MQRPA & FAM & MQRPA   & FAM-A     
                    & FAM-B   & FAM-C     
                    & FAM-D    \\ \hline
1.3185 & 1.3183 & 1.557(-3) & 1.547(-3) & 1.547(-3) & 1.547(-3) & 1.547(-3) \\
1.3731 & 1.3731 & 5.771(-5) & 5.810(-5) & 5.818(-5) & 5.827(-5) & 5.824(-5) \\
2.4582 & 2.4581 & 1.968(-6) & 1.643(-6) & 1.896(-6) & 2.188(-6) & 2.047(-6) \\
2.5998 & 2.5975 & 8.978(-5) & 8.870(-5) & 8.894(-5) & 8.919(-5) & 8.907(-5) \\
3.6687 & 3.6657 & 1.555(-5) & 8.681(-6) & 1.140(-5) & 1.498(-5) & 1.515(-5) \\
5.1185 & 5.1212 & 3.907(-2) & 3.885(-2) & 3.899(-2) & 3.914(-2) & 3.914(-2) \\
7.4108 & 7.4084 & 1.388(-5) & 2.926(-5) & 2.228(-5) & 1.697(-5) & 1.622(-5) \\
\end{tabular}
\end{ruledtabular}
\end{table*}
%%%%%%%%%%%%%%%%%%%%%%%%%%%%%%%%%%%%%%%%%%%%%%%%%%%%%%%%%%%%%%%%%%%%%
% Tables I and II end
%%%%%%%%%%%%%%%%%%%%%%%%%%%%%%%%%%%%%%%%%%%%%%%%%%%%%%%%%%%%%%%%%%%%%

In Tables~\ref{table:24MgIS} and \ref{table:24MgIV} we show the energies and transition strength of the low-lying $K=0$ QRPA modes, calculated with  MQRPA and FAM.
Although the centers of the contours in the complex $\omega_\gamma$-plane are only approximately chosen, the low-lying QRPA eigenfrequencies calculated with  FAM are in good agreement with  MQRPA results.
The agreement is excellent  for the lowest-lying states, which are usually of more interest. The FAM  transition strength was obtained  using the  techniques A-D described in Sec.~\ref{sec:famdiscrete}. It is gratifying 
to see that the four methods generally agree at least up to two decimal places, except for the modes carrying very  small strength  ($\sim 10^{-4}\,e^2$fm$^4$).
The nice agreement between FAM-C and FAM-D  results demonstrates  the consistency between the two sets of QRPA amplitudes calculated from the isoscalar 
and isovector  external monopole fields. 
The difference between the strengths obtained by the MQRPA and FAM is consistent with the convergence criteria used in the FAM iterations.

%%%%%%%%%%%%%%%%%%%%%%%%%%%%%%%%%%%%%%%%%%%%%%%%%%%%%%%%%%%%%%%%%%%%%
% Fig. 2
%%%%%%%%%%%%%%%%%%%%%%%%%%%%%%%%%%%%%%%%%%%%%%%%%%%%%%%%%%%%%%%%%%%%%
\begin{figure}[htbp]
\includegraphics[width=0.9\columnwidth]{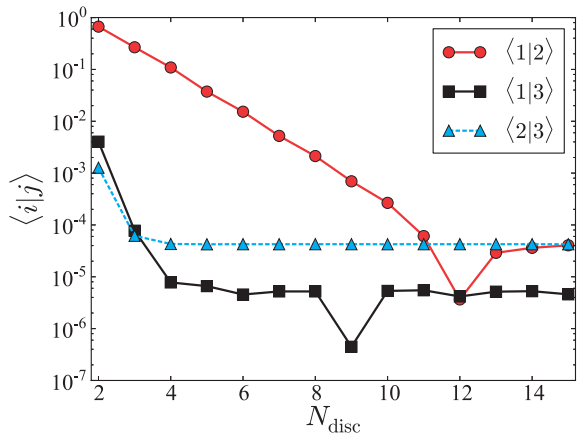}
\caption{(Color online) 
Convergence of the orthogonality of the states as a number of discretization points of a circular contour with radius 0.02\,MeV.
The first three low-lying states labeled as 1 (1.32\,MeV), 2 (1.37\,MeV), and 3 (2.46\,MeV) are shown.\label{fig:convergence}}
\end{figure}

Figure~\ref{fig:convergence} demonstrates the convergence of the QRPA amplitudes
against the number of discretization points $N_{\rm disc}$ used in the contour integration. Specifically, it shows 
the orthogonality of the QRPA amplitudes for the  three lowest QRPA states.
The orthogonality between the first and third state,  and between the second and third state,  is
 achieved already at $N_{\rm disc}=4$, while the convergence for  the pair of first and second states  is slower. This is not surprising as
the energies of the first and second QRPA roots differs only by 0.05\,MeV; hence,
and a finer integration mesh is required  to remove the contribution from the unwanted pole outside of the contour.
Our  results show that  to obtain the convergence for the contour integration, consistent with the accuracy required during the regular FAM iterations,
taking 11 points is fully sufficient, at least for the two lowest  modes.

Lastly, we discuss an example of an incorrect contour assignment following the discussion in Sec.~\ref{sec:famdiscrete}.
A contour of radius of 0.2\,MeV, centered at 1.3\,MeV, includes the first two QRPA solutions at 1.32\,MeV and 1.37\,MeV.
Such an incorrect choice would be made  if the contour were determined
from the isoscalar strength function calculated with a resolution of $\gamma=0.02$\,MeV  shown in Fig.~\ref{fig:24Mg-ISM}.
The calculated isoscalar (isovector) monopole transition strength determined
according to A-C is $1.57\times 10^{-2}$ ($1.61\times 10^{-3}$) $e^2\,{\rm fm}^4$, which is  precisely the sum of the two QRPA strengths.
However, method D completely fails, yielding the  values of $6\times 10^{-8}\,e^2\,{\rm fm}^4$ (isoscalar) and 
$3\times 10^{-9}\,e^2\,{\rm fm}^4$ (isovector), that are clearly off from those obtained with  the procedures A-C.
The QRPA frequency calculated from Eq.~(\ref{eq:omega2}) with the isoscalar (isovector) monopole external field is 1.371\,MeV (1.320\,MeV).
Since the isoscalar (isovector) monopole strength of the second (first) QRPA state is larger than that of the first (second) QRPA state
by two orders of magnitude, the QRPA frequency calculated using the contour, which encloses both  poles, is close to  the energy of the collective state.
Therefore, the consistency between the results obtained in methods C and D, together with the value  of  weighted frequency, can be used to find the contour that encloses a single QRPA pole.

%%%%%%%%%%%%%%%%%%%%%%%%%%%%%%%%%%%%%%%%%%%%%%%%%%%%%%%%%%%%%%%%%%%%%%%%%%%%%%%%%%%%%%%%%%%%%%%%%%%%%%%%%%%%%%%%%%%%%%%%
%
%  Numerical calculations, rare-earths
%
%%%%%%%%%%%%%%%%%%%%%%%%%%%%%%%%%%%%%%%%%%%%%%%%%%%%%%%%%%%%%%%%%%%%%%%%%%%%%%%%%%%%%%%%%%%%%%%%%%%%%%%%%%%%%%%%%%%%%%%%
\subsection{Low-lying QRPA modes in deformed rare-earth nuclei}

To demonstrate the feasibility of the  FAM-QRPA formalism to describe the low-lying collective modes of deformed  nuclei,
we have performed FAM calculations for the low-lying $K=0$ strength of
$^{166,168,172}$Yb, and $^{170}$Er, which were previously studied with  MQRPA  in Refs.~\cite{PhysRevC.82.034326,PhysRevC.84.014332}.
The calculations were carried out using SkM*  Skyrme EDF \cite{Bartel198279} with the volume pairing. The pairing strengths have been adjusted
to reproduce the
odd-even binding energy difference in  $^{172}$Yb evaluated  with the three-point expression. They are:
$V_{\rm n}=-176\,{\rm MeV\,fm}^{-3}$ and $V_{\rm p}=-218\,{\rm MeV\,fm}^{-3}$ for the  quasiparticle energy cutoff $E_{\rm cut}=60$\,MeV and 
$V_{\rm n}=-150\,{\rm MeV\,fm}^{-3}$ and $V_{\rm p}=-177.5\,{\rm MeV\,fm}^{-3}$ for $E_{\rm cut}=200$\,MeV. 
To obtain QRPA amplitudes in FAM, we applied the isoscalar quadrupole $K=0$ external field~\cite{PhysRevC.71.034310}, and 
the electric reduced matrix elements $B(E2)$ for the excitational modes discussed in Refs.~\cite{PhysRevC.82.034326,PhysRevC.84.014332} 
are computed using the technique D described in Sec.~\ref{sec:famdiscrete}.
The transformed harmonic oscillator  basis with 20 major oscillator shells was employed.  To compute residua, we used the circular contours 
 with radii 0.1\,MeV  for $^{166}$Yb, $^{172}$Yb, and $^{170}$Er,
and with radii 0.01\,MeV for $^{168}$Yb. 
The locations of the contour centers  estimated from the conventional FAM calculations are:
1.40\,MeV, 1.75\,MeV, 1.30\,MeV, and 1.30\,MeV for $^{166}$Yb, $^{168}$Yb, $^{172}$Yb, and $^{170}$Er, respectively. 
The results were compared with the MQRPA calculations of Refs.~\cite{PhysRevC.82.034326,PhysRevC.84.014332} employing a different HFB solver and  an additional cutoff associated 
with the occupation probabilities of canonical states.

%%%%%%%%%%%%%%%%%%%%%%%%%%%%%%%%%%%%%%%%%%%%%%%%%%%%%%%%%%%%%%%%%%%%%
% Table III 
%%%%%%%%%%%%%%%%%%%%%%%%%%%%%%%%%%%%%%%%%%%%%%%%%%%%%%%%%%%%%%%%%%%%%
\begin{table}[htbp]
\begin{threeparttable}[b]
\caption{FAM-QRPA energies and  $B(E2)$ values of the low-lying $K=0$ states in $^{166}$Yb, $^{168}$Yb, $^{172}$Yb, and $^{170}$Er at $E_{\rm cut}=200$\,MeV compared to the MQRPA results of
Ref.~\cite{PhysRevC.84.014332}. The additional result for $^{172}$Yb
corresponding $E_{\rm cut}=60$\,MeV is compared to the MQRPA values obtained in Ref.~\cite{PhysRevC.82.034326}.
\label{table:Yb2}}
\begin{ruledtabular}
\begin{tabular}{ccccc}
\multirow{2}{*}{nucleus}
 & \multicolumn{2}{c}{$\Omega_i$ (MeV)}  & \multicolumn{2}{c}{$B(E2)$ ($e^2$b$^2$)} \\
          & MQRPA      & FAM     & MQRPA        & FAM   \\
 \hline \\[-6pt]
$^{166}$Yb & 1.802      & 1.422   & 0.0398      & 0.0327 \\ 
$^{168}$Yb & 2.039      & 1.747   & 0.0343      & 0.0186 \\ 
$^{172}$Yb & 1.605      & 1.306   & 0.0049      & 0.0088 \\
$^{170}$Er & 1.596      & 1.322   & 0.0030      & 0.0047 \\ [4pt]
$^{172}$Yb$^a$ & 1.390      & 1.319   & 0.0050      & 0.0092    
\end{tabular}
\end{ruledtabular}
 \begin{flushleft}
$^a$$E_{\rm cut}=60$\,MeV
 \end{flushleft}
\end{threeparttable}
\end{table}
%%%%%%%%%%%%%%%%%%%%%%%%%%%%%%%%%%%%%%%%%%%%%%%%%%%%%%%%%%%%%%%%%%%%%
Table~\ref{table:Yb2} displays the results for excitation energies and $B(E2)$ rates  of  the $K=0$ QRPA modes.
For $^{172}$Yb the calculations were carried out with two quasiparticle cutoffs:
$E_{\rm cut}=60$\,MeV and 200\,MeV. 
The  FAM-QRPA excitation energy is close to the MQRPA value
with $E_{\rm cut}=60$\,MeV~\cite{PhysRevC.82.034326}, but this agreement does not hold when $E_{\rm cut}$ is increased. Indeed, 
while our FAM-QRPA values weakly depend on $E_{\rm cut}$,
a 15\% increase of the MQRPA energy for $^{172}$Yb was reported in 
Ref.~\cite{PhysRevC.84.014332} when going to $E_{\rm cut}=200$\,MeV.
Interestingly, the  $B(E2)$ values weakly depend on energy cutoff in both methods. However, the $B(E2)$ values obtained in FAM-QRPA  are 
twice as large as the MQRPA results.

For $^{166}$Yb, $^{168}$Yb, and $^{170}$Er,
the excitation energies obtained in  MQRPA  are larger by 0.3-0.4\,MeV than those in FAM-QRPA.
The agreement between  $B(E2)$ values is good in $^{166}$Yb, but 
gets worse in the other cases studied. It is difficult to speculate what is the origin of those differences. We note, however, that (i) the HFB solvers used in both calculations are different (see benchmarking results in Ref.~\cite{Pei08}), and (ii) there are additional canonical energy cutoffs in  in  MQRPA \cite{PhysRevC.71.034310} that are not present in our FAM-QRPA method.

%%%%%%%%%%%%%%%%%%%%%%%%%%%%%%%%%%%%%%%%%%%%%%%%%%%%%%%%%%%%%%%%%%%%%
% Table IV
%%%%%%%%%%%%%%%%%%%%%%%%%%%%%%%%%%%%%%%%%%%%%%%%%%%%%%%%%%%%%%%%%%%%%
\begin{table}[htbp]
\caption{
Isoscalar and isovector quadrupole strength (in $e^2\,{\rm fm}^4$) of the low-lying $K=0$ states in  $^{166}$Yb, $^{168}$Yb, $^{172}$Yb, and $^{170}$Er shown in
 Table~\ref{table:Yb2} with $E_{\rm cut}=200$\,MeV. The isoscalar (isovector) strength in FAM-C is calculated 
using the isoscalar (isovector) quadrupole external field. The isoscalar (isovector) 
quadrupole strength  in FAM-D is calculated using the QRPA amplitudes obtained from the FAM calculation using the isovector (isoscalar) 
quadrupole external field. \label{table:ISQIVQ}}
\begin{ruledtabular}
\begin{tabular}{ccccc}
\multirow{2}{*}{nucleus} & \multicolumn{2}{c}{ISQ} & \multicolumn{2}{c}{IVQ} \\
          & FAM-C    & FAM-D  & FAM-C      & FAM-D   \\ \hline \\[-6pt]
$^{166}$Yb & 299.854 & 299.856 & 0.585519    & 0.585520 \\
$^{168}$Yb & 160.126 & 160.127 & 0.969114    & 0.969124 \\
$^{172}$Yb & 93.2710 & 93.2735 & 0.081406    & 0.081404 \\
$^{170}$Er & 56.2932 & 56.2913 & 0.460285    & 0.460254 \\
\end{tabular}
\end{ruledtabular}
\end{table}
%%%%%%%%%%%%%%%%%%%%%%%%%%%%%%%%%%%%%%%%%%%%%%%%%%%%%%%%%%%%%%%%%%%%%

The isoscalar 
quadrupole (ISQ) and isovector quadrupole (IVQ) strengths
are displayed 
in Table~\ref{table:ISQIVQ} for the deformed nuclei shown in
 Table~\ref{table:Yb2}.
The strengths obtained from the QRPA amplitudes derived from two external quadrupole fields agree excellently.
This result  clearly shows that the contours used in these calculations enclose only a single QRPA pole.

%%%%%%%%%%%%%%%%%%%%%%%%%%%%%%%%%%%%%%%%%%%%%%%%%%%%%%%%%%%%%%%%%%%%%%%%%%%%%%%%%%%%%%%%%%%%%%%%%%%%%%%%%%%%%%%%%%%%%%%%
%
%  Conclusions
%
%%%%%%%%%%%%%%%%%%%%%%%%%%%%%%%%%%%%%%%%%%%%%%%%%%%%%%%%%%%%%%%%%%%%%%%%%%%%%%%%%%%%%%%%%%%%%%%%%%%%%%%%%%%%%%%%%%%%%%%%
\section{Conclusions} \label{sec:conclusion}

We have formulated  and tested the FAM-QRPA method for efficient computations of  discrete QRPA modes. The new framework is based on the application of Cauchy's integral formula to the FAM amplitudes defined in the complex frequency plane. The method is fully self-consistent and does not require any configuration-space  truncations at the QRPA level.
The method is  particularly useful when applied to  the isolated collective  QRPA modes.
For the description of the transition strength carried by densely distributed modes, the conventional FAM formulation  is more 
appropriate.

The FAM-QRPA method has been benchmarked and tested  
by comparing with MQRPA results for an oblate configuration of
$^{24}$Mg.
Illustrative examples of  large-scale  calculations have been presented for the $K=0$ isoscalar and isovector quadrupole modes of 
deformed rare-earth nuclei $^{166}$Yb, $^{168}$Yb, $^{172}$Yb, and $^{170}$Er.

Our results demonstrate that the proposed  formulation of  FAM-QRPA can be used as an efficient tool to calculate discrete QRPA modes  of heavy, deformed, and  superfluid nuclei. Once the contour around the mode of interest  is specified, 
the FAM-QRPA method allows one to perform a fully self-consistent QRPA calculation employing {\it the same}  model space  as in  HFB. 
Thanks to the rapid convergence achieved with  Broyden's method used in our implementation, FAM-QRPA is amenable to high-performance parallel computing. This offers promise of systematic calculations of various kinds of low-lying excitations and decays over the entire nuclear landscape. Of particular importance are QRPA studies of low-energy dipole and quadrupole states, $\beta$ decays, and $\beta\beta$ decays. 
The work on extending the FAM-QRPA formalism to  $K\ne 0$ and charge-exchange modes is in progress.

%%%%%%%%%%%%%%%%%%%%%%%%%%%%%%%%%%%%%%%%%%%%%%%%%%%%%%%%%%%%%%%%%%%%%%%%%%%%%%%%%%%%%%%%%%%%%%%%%%%%%%%%%%%%%%%%%%%%%%%%
%
%  Acknowledgments
%
%%%%%%%%%%%%%%%%%%%%%%%%%%%%%%%%%%%%%%%%%%%%%%%%%%%%%%%%%%%%%%%%%%%%%%%%%%%%%%%%%%%%%%%%%%%%%%%%%%%%%%%%%%%%%%%%%%%%%%%%
\begin{acknowledgments}
Useful discussions with J. Dobaczewski and T. Nakatsukasa
are gratefully acknowledged. This work was
supported by
the U.S. Department of Energy under
Contract Nos.\ DE-FG02-96ER40963
 (University of Tennessee), 
DE-SC0008499    (NUCLEI SciDAC Collaboration), 
by JUSTIPEN (Japan-U.S. Theory Institute for Physics with Exotic Nuclei) under grant
number No.\ DEFG02-06ER41407 (University of Tennessee),
by the Academy of Finland under the Centre of Excellence Programme 2012--2017
(Nuclear and Accelerator Based Physics Programme at JYFL), and FIDIPRO programme.  An award of computer time was provided by the Innovative and Novel Computational Impact on Theory and Experiment (INCITE) program. 
\end{acknowledgments}

\bibliographystyle{apsrev4-1}
\bibliography{discretefam.bib}% Produces the bibliography via BibTeX.

\end{document}